\documentclass[a4paper,fleqn,usenatbib]{mnras}
\usepackage{txfonts}

\usepackage{ae,aecompl}
\usepackage{graphicx}	

\title[GAMA: 1.4GHz Radio SFR]{Galaxy And Mass Assembly: The 1.4\,GHz SFR indicator, SFR-M$_{*}$ relation and predictions for ASKAP-GAMA }
\author[L. J. M. Davies et. al.]{L. J. M. Davies$^{1}$\thanks{E-mail:
 luke.j.davies@uwa.edu.au}, M. T. Huynh$^{1}$, A. M. Hopkins$^{2}$, N. Seymour$^{3}$, S. P. Driver$^{1,4}$, \newauthor A. G. R. Robotham$^{1}$, I. K. Baldry$^{5}$, J. Bland-Hawthorn$^{6}$, N. Bourne$^{7}$, M. N. Bremer$^{8}$,  \newauthor   M. J. I. Brown$^{9}$,  S. Brough$^{2}$, M. Cluver$^{10}$, M. W. Grootes$^{11}$, M. Jarvis$^{12,13}$, J. Loveday$^{14}$,  \newauthor A. Moffet$^{1}$,  M. Owers$^{2,15}$, S. Phillipps$^{}$,  E. Sadler$^{6}$,   L. Wang$^{16,17}$, S. Wilkins$^{14}$,   \newauthor A. Wright$^{1}$ \\
 \\
$^{1}$ ICRAR, The University of Western Australia, 35 Stirling Highway, Crawley, WA 6009, Australia\\
$^{2}$ Australian Astronomical Observatory, P.O. Box 915, North Ryde,  NSW 1670, Australia\\
$^{3}$ ICRAR, Curtin University, Bentley, WA 6102, Australia\\
$^{4}$ SUPA, School of Physics and Astronomy, University of St Andrews, North Haugh, St Andrews, Fife, KY16 9SS, UK\\
$^{5}$ Astrophysics Research Institute, Liverpool John Moores University, IC2, Liverpool Science Park, 146 Brownlow Hill, Liverpool, L3 5RF, UK\\
$^{6}$ Sydney Institute for Astronomy, School of Physics A28, University of Sydney, NSW 2006, Australia.\\
$^{7}$ Institute for Astronomy, University of Edinburgh, Royal Observatory, Blackford Hill, Edinburgh EH9 3HJ, UK \\
$^{8}$ Astrophysics Group, School of Physics, University of Bristol, Tyndall Avenue, Bristol BS8 1TL\\
$^{9}$ School of Physics and Astronomy, Monash University, Clayton, Victoria 3800, Australia \\
$^{10}$ Department of Physics and Astronomy, University of the Western Cape, Robert Sobukwe Road, Bellville, 7535, South Africa\\
$^{11}$ ESA/ESTEC, 2201 AZ Noordwijk, The Netherlands\\
$^{12}$ Astrophysics, University of Oxford, Denys Wilkinson Building, Keble Road, Oxford, OX1 3RH, UK \\
$^{13}$ Department of Physics, University of the Western Cape, Bellville 7535, South Africa \\
$^{14}$ Astronomy Centre, University of Sussex, Falmer, Brighton BN1 9QH, UK \\
$^{15}$ Department of Physics and Astronomy, Macquarie University, NSW 2109, Australia \\
$^{16}$ SRON Netherlands Institute for Space Research, Landleven 12, 9747 AD, Groningen, The Netherlands \\
$^{17}$ Kapteyn Astronomical Institute, University of Groningen, Postbus 800, 9700 AV Groningen, the Netherlands}

\date{Accepted XXX. Received YYY; in original form ZZZ}

\pubyear{2016}

\begin{document}
\label{firstpage}
\pagerange{\pageref{firstpage}--\pageref{lastpage}}
\maketitle

\begin{abstract}

We present a robust calibration of the 1.4\,GHz radio continuum star formation rate (SFR) using a combination of the Galaxy And Mass Assembly (GAMA) survey and the Faint Images of the Radio Sky at Twenty-cm (FIRST) survey. We identify individually detected 1.4\,GHz GAMA-FIRST sources and use a late-type, non-AGN, volume-limited sample from GAMA to produce stellar mass-selected samples. The latter are then combined to produce FIRST-stacked images. This extends the robust parametrisation of the 1.4\,GHz-SFR relation to faint luminosities. For both the individually detected galaxies and our stacked samples, we compare 1.4\,GHz luminosity to SFRs derived from GAMA to determine a new 1.4\,GHz luminosity-to-SFR relation with well constrained slope and normalisation. For the first time, we produce the radio SFR-M$_{*}$ relation over 2\,decades in stellar mass, and find that our new calibration is robust, and produces a SFR-M$_{*}$ relation which is consistent with all other GAMA SFR methods. Finally, using our new 1.4\,GHz luminosity-to-SFR calibration we make predictions for the number of star-forming GAMA sources which are likely to be detected in the upcoming ASKAP surveys, EMU and DINGO.

\end{abstract}

\begin{keywords}
galaxies: star-formation - galaxies: evolution - radiation mechanisms: non-thermal - radio continuum: galaxies
\end{keywords}

\section{Introduction}

The rate at which galaxies are forming new stars (the star formation rate, SFR) is critical to our understanding of the formation of stellar mass in galaxies and the global evolution of baryonic matter in the Universe. However, accurately measuring SFRs is problematic. This is largely due to the fact that common methods for deriving SFR are limited by either dust obscuration \citep[$e.g.$ see][]{Meurer99} and/or aperture corrections to account for missing flux in fibre-based spectroscopy \citep[for example using the H$\alpha$ emission line to derive SFRs, $e.g.$][]{Hopkins13,Gunawardhana13}. 

A potentially more robust approach is to measure both the UV and total IR emission simultaneously \citep[UV+TIR or full spectral energy distribution (SED)-derived SFRs $e.g.$][]{Bell05,Papovich07,Barro11}, probing both the dust obscured and unobscured SFRs. This approach does not require obscuration corrections, as one completely observes the full (direct and reprocessed) emission from young stars. The number of sources with robust UV+TIR measurements however, has historically been very small, hampering efforts to analyse large samples of galaxies using this method.   

Recently great strides have been made in improving techniques to derive robust SFRs for large samples of galaxies \cite[see][hereafter D16]{Davies16b}. Complex prescriptions for the treatment of obscuration corrections in the UV, such as using radiative transfer (RT) models \cite[][Grootes in prep and D16]{Tuffs04,Wood08,Popescu11,Popescu13,Grootes13,Grootes14}, have dramatically improved our ability to reduce the scatter in UV derived SFRs to that of the intrinsic population. Furthermore, samples of UV+TIR detected sources have increased dramatically with the extensive surveys of GALEX \citep{Martin05} and \textit{Herschel} \citep{Pilbratt10}, and improvements to SED modelling, such as \textsc{magphys} \citep{daCunha08} and \textsc{cigale} \citep[$e.g.$][]{Noll09}, have allowed us to probe statistically robust samples using UV+TIR SFRs \citep[for example see D16 and][]{SmithD12}. 

Despite these improvements, it is also possible to avoid sources of error induced by obscuration corrections and aperture corrections by using a measure of star-formation which is unaffected by dust obscuration, integrated over the whole galaxy and probing down to faint levels. The radio continuum is ideally suited to this. It has long been known that there is a tight correlation between FIR emission and rest-frame 1.4\,GHz radio power \citep[$e.g.$][]{Kruit71, Helou85, Condon92, Yun01}. This relation arises because emission at both wavelengths is connected with ongoing star formation. Emission from star forming galaxies at 1.4\,GHz is dominated by synchrotron radiation arising from relativistic electrons thought to be accelerated by supernovae shocks \citep[$e.g.$][]{Harwit75}. Given that massive stars dominate both the supernova rate and dust heating, the FIR-radio correlation arrises through the same underlying sources producing the emission at both wavelengths. As the supernova rate is intimately linked to the birth of high mass stars and emission at these wavelengths is unencumbered by dust obscuration, the non-thermal radio luminosity provides a robust and dust-insensitive measure of the current star-formation on $\sim100$\,Myr timescales \citep[$e.g.$ see][]{Condon02}. Thermal radio emission is also strongly correlated with star-formation \cite[$e.g.$][]{Galvin16}, but has a different spectral slope to non-thermal emission \citep[$e.g.$][]{Condon92}, and in this work we assume that that thermal contribution at 1.4\,GHz is negligible \citep[as found for the majority of local star-forming galaxies,][]{Rabidoux14}.

In order to robustly use 1.4\,GHz emission to probe dust-unbiased star-formation, we require the observed 1.4\,GHz radio power to be well calibrated against reliable measures of star-formation using other tracers. There are two different approaches to perform such a calibration: (i) using detailed observations of well studied nearby galaxies in the high signal to noise regime, primarily with dedicated observations \citep[$e.g.$][]{Kennicutt09, Murphy11, Heesen14}, and (ii) the statistical approach of identifying multiple faint sources in large area surveys \citep[$e.g.$][]{Bell03,Hopkins03}. Until recently, the latter approach has relied on SFR tracers that require dust obscuration and/or aperture corrections to calibrate the 1.4\,GHz SFR indicator and have been limited to relatively high radio luminosity systems. With the new full SED and well calibrated SFR measures from surveys such as the Galaxy And Mass Assembly \citep[GAMA][]{Driver11, Liske15} as in D16 however, we can begin to explore the 1.4\,GHz SFR indicator without the need for complex corrections and to significantly lower radio luminosities. In this work we utilise the UV+TIR and \textsc{magphys}-derived SFRs from D16 (these are described briefly in Section 4) to produce a new 1.4\,GHz luminosity to SFR calibration and use stacking techniques to extend the 1.4\,GHz luminosity - SFR relation to faint luminosities.

Such calibrations will become extremely powerful with the next generation of deep large area radio continuum surveys from the Square Kilometre Array (SKA) and its precursors, such as ASKAP-EMU \citep{Norris11} and MeerKAT-MIGHTEE \citep{Jarvis12}. In preparation for these future studies, it is essential that we fully exploit existing datasets in order to explore SFRs derived from the 1.4\,GHz radio emission. Here we use the current state of the art large area radio survey, FIRST, in combination with GAMA to investigate the 1.4\,GHz SFR indicator and make predictions for number of GAMA sources what will be detectable with ASKAP. Throughout this paper we use a standard $\Lambda$CDM cosmology with {H}$_{0}$\,=\,70\,kms$^{-1}$\,Mpc$^{-1}$, $\Omega_{\Lambda}$\,=\,0.7 and $\Omega_{\mathrm{M}}$\,=\,0.3.

\section{Data}

\subsection{GAMA}

The extended GAMA survey (GAMA II) covers 286\,deg$^{2}$ to a main survey limit of $r_{\mathrm{AB}}<19.8$\,mag in three equatorial regions  (G09, G12 and G15) and two southern regions (G02 and G23 \- survey limit of $i_{\mathrm{AB}}<19.2$\,mag in G23) \citep{Liske15}. The limiting magnitude of GAMA was initially designed to probe all aspects of cosmic structures on 1\,kpc to 1\,Mpc scales spanning all environments and out to a redshift limit of $z\sim$0.4. The spectroscopic survey was undertaken using the AAOmega fibre-fed spectrograph \citep[][]{Sharp06,Saunders04} in conjunction with the Two-degree Field \citep[2dF,][]{Lewis02} positioner on the Anglo-Australian Telescope and obtained redshifts for $\sim$280,000 targets covering $0<z\lesssim0.5$ with a median redshift of $z\sim0.2$, and highly uniform spatial completeness \citep[see][for summaries of GAMA observations]{Baldry10,Robotham10,Driver11}. 

\begin{figure}
\begin{center}
\includegraphics[scale=0.6]{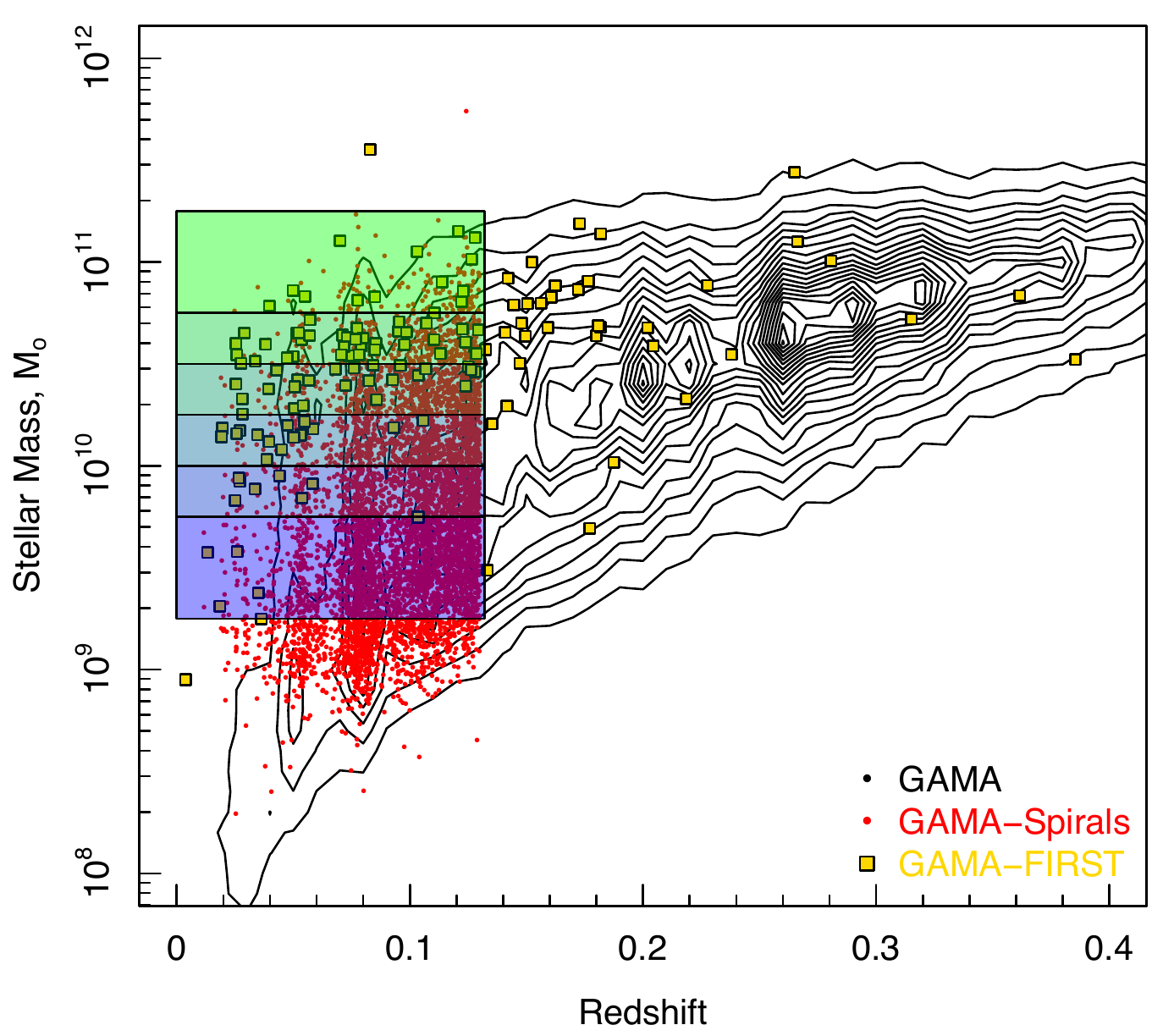}
\end{center}
\caption{The redshift-M$_{*}$ distribution of GAMA II$_{Eq}$ galaxies (contours), the GAMA volume limited spirals sample of \citet{Grootes14} used in our stacking analysis (red circles), and the final star-forming GAMA-FIRST sample, excluding all potential AGN sources (gold squares). The coloured shaded regions display the volume stacked within carefully designed stellar mass bins.}
\label{fig:Mz}
\end{figure}

Full details of the GAMA survey can be found in \citet{Hopkins13, Driver11, Driver16} and \citet{Liske15}. In this work we use the data obtained in the three equatorial regions, which we refer to here as GAMA II$_{Eq}$. Stellar masses for the GAMA II$_{Eq}$ sample are derived from the $ugriZYJHK$ photometry using a method similar to that outlined in \cite{Taylor11} assuming a Chabrier IMF \citep{Chabrier03}. Figure \ref{fig:Mz} displays the stellar mass-redshift distribution of the GAMA II$_{Eq}$ sample. All photometry used in this work comes from the \textsc{lambdar} catalogue discussed in \cite{Wright16} and spectral line analysis will be detailed in Gordon et al (in prep).

\subsection{FIRST}

The Faint Images of the Radio Sky at Twenty-cm (FIRST) survey \citep{Becker95} is a 1.4\,GHz continuum survey in the Northern hemisphere and contains $\sim$90 sources deg$^{-2}$ at the 1\,mJy survey threshold to an rms sensitivity of $\sim$0.15\,mJy\,beam$^{-1}$. The survey was undertaken by the VLA in B configuration with a synthesized restoring beam of 5.4$^{\prime\prime}$ full width at half-maximum. We use the `14Dec17' FIRST catalogue which contains observations from 1993 to 2011. This catalogue consists of  946,432 sources covering $\sim$10,500 deg$^{2}$ ($i.e. \sim$95\,deg$^{-2}$).

\section{Combining GAMA and FIRST}

\subsection{GAMA-FIRST Detected Sample}
\label{sec:sel} 

To identify GAMA galaxies which have a detection in FIRST, we perform a 3$^{\prime\prime}$ cross match \citep[comparable to the FIRST half beam width, see similar crossmatching in $e.g.$][]{Sadler07} between the GAMA II$_{Eq}$ galaxies with robust redshifts and photometric measurements, and the FIRST catalogues. Where multiple GAMA sources are matched to a single FIRST detection ($<2\%$ of sources), we assign the closest position match. This results in 1991 matched galaxies in the GAMA volume, which we refer to as the GAMA-FIRST sample. We highlight that this sample is comparable to the sample obtained by Ching et al. (in press) who perform a more complex match between SDSS and FIRST galaxies in the GAMA regions.

\begin{figure*}
\begin{center}
\includegraphics[scale=0.55]{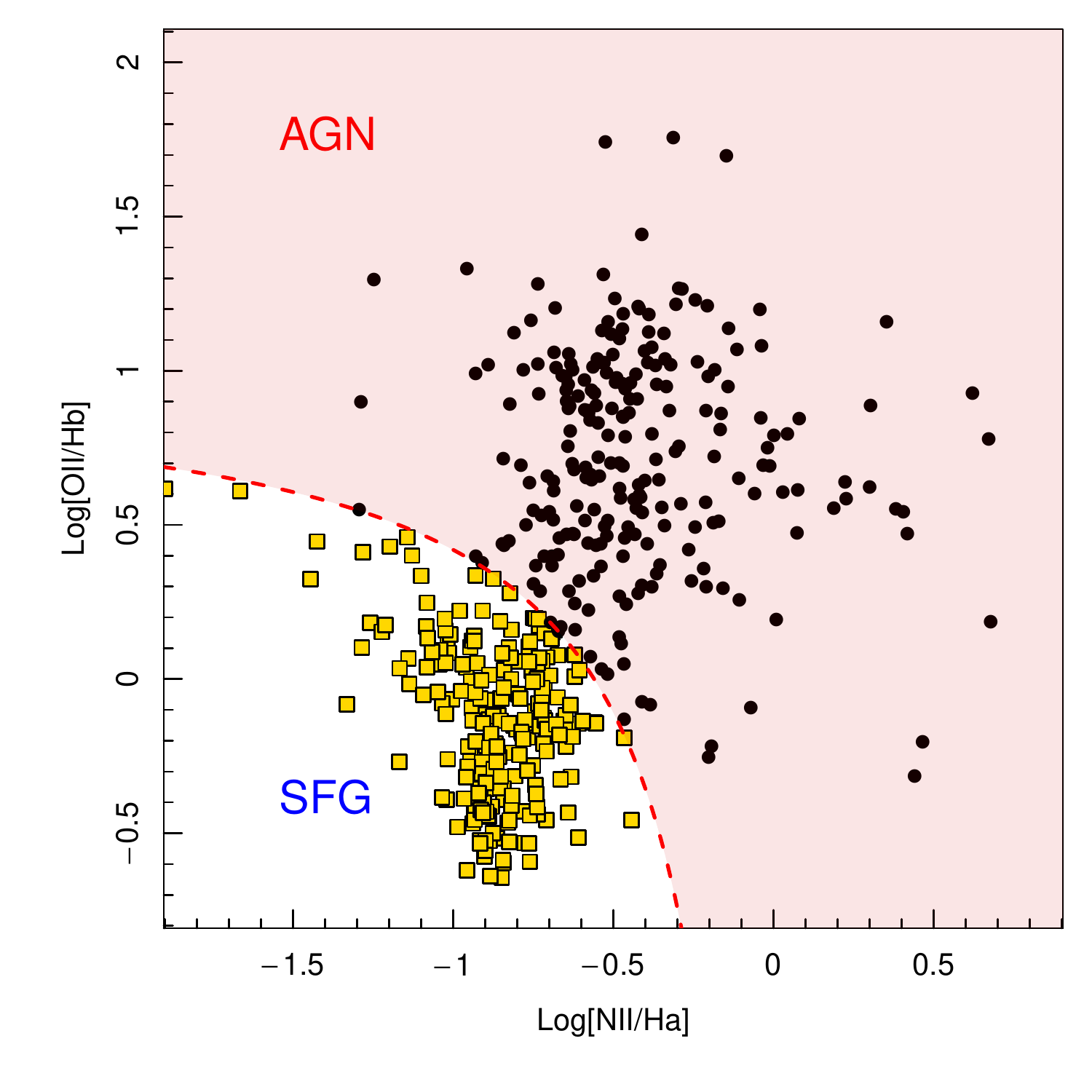}
\includegraphics[scale=0.55]{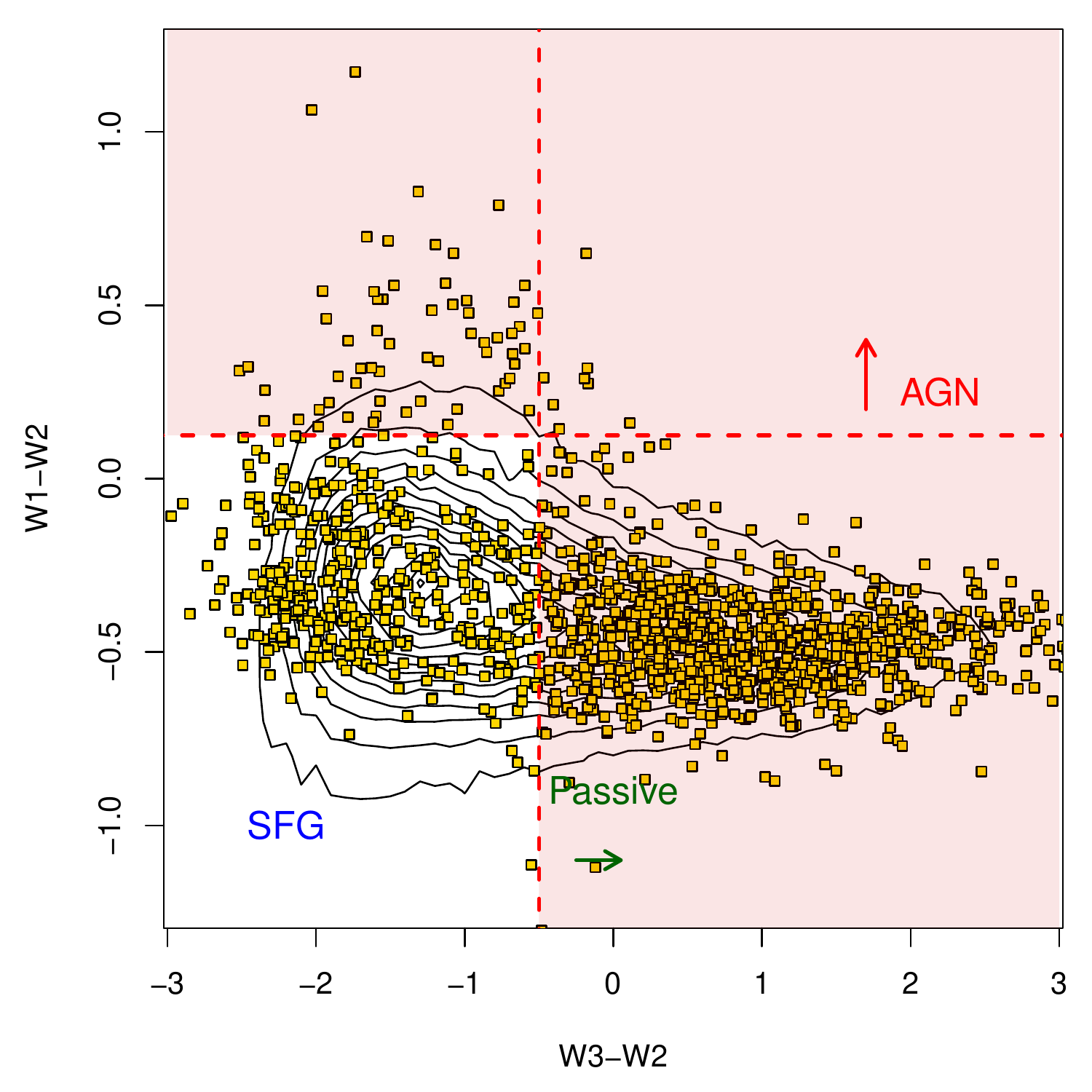}
\vspace{-5mm}
\includegraphics[scale=0.55]{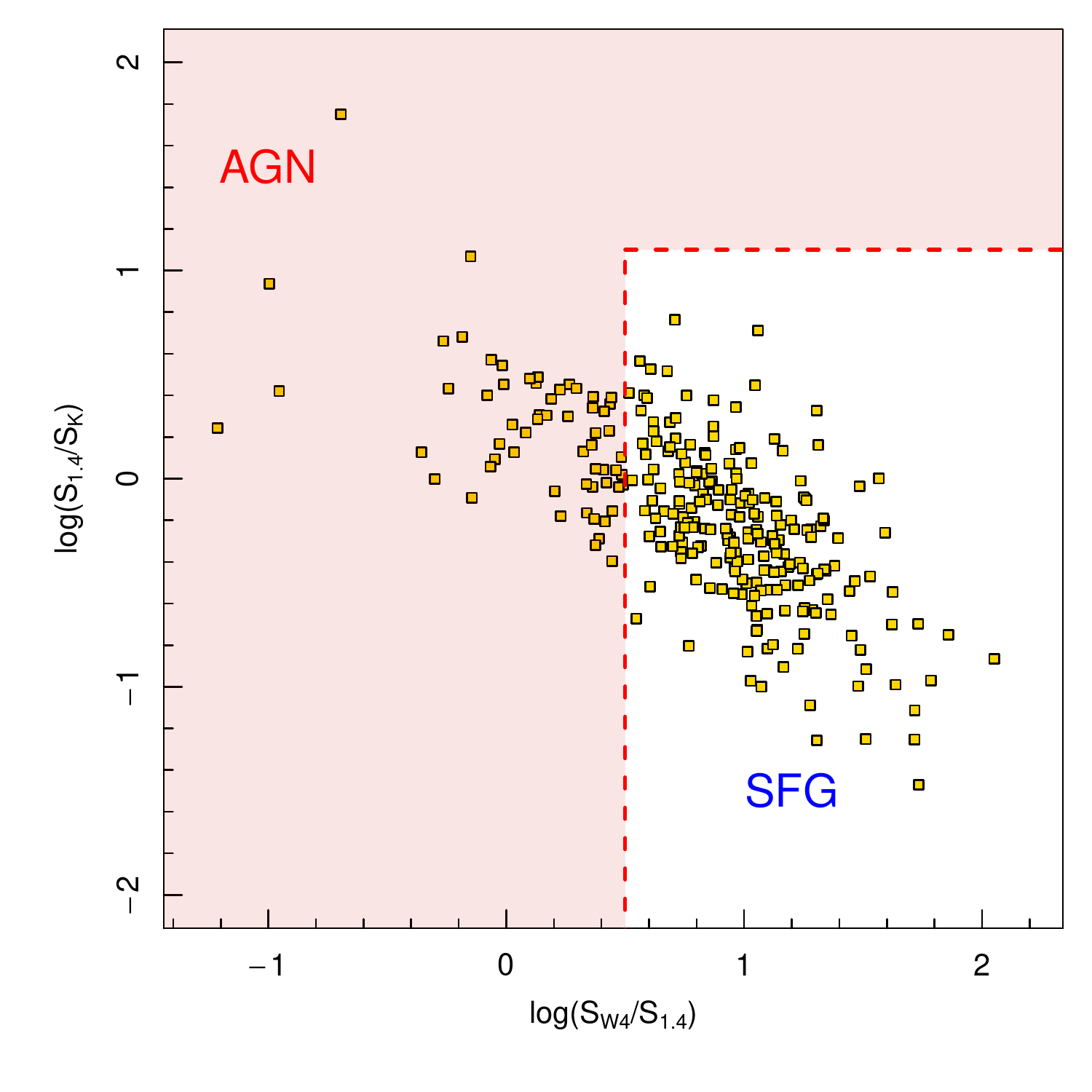}
\includegraphics[scale=0.55]{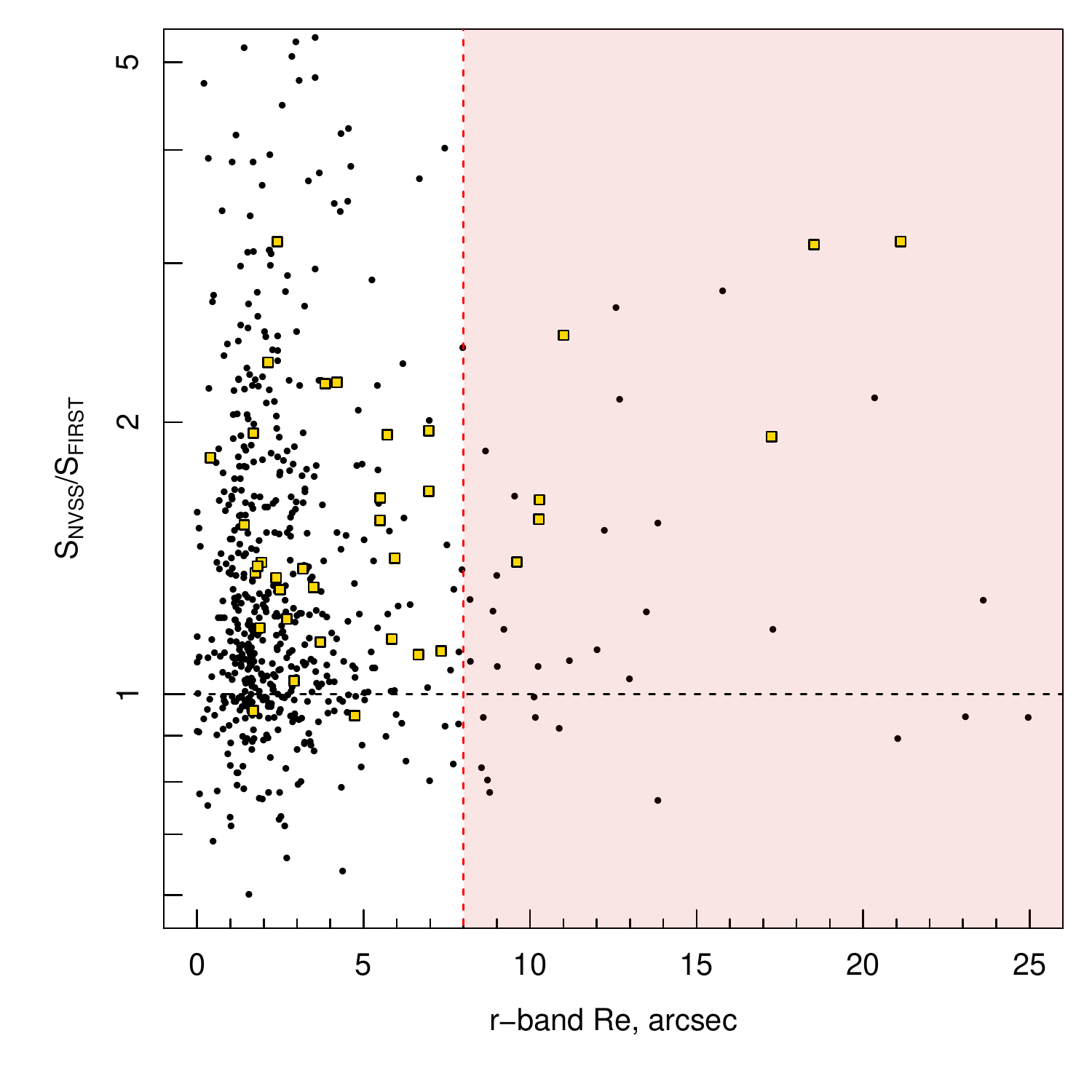}

\end{center}
\caption{Identification of star-forming galaxies in GAMA-FIRST sample, regions where sources are excluded in our sample selection are shaded red. Top left: BPT classification used to select star-forming galaxies from the GAMA-FIRST crossmatched sample. Points display GAMA-FIRST matched sources which are detected at $>2\sigma$ in all emission lines required for the BPT. The red dashed line displays the SF-AGN dividing line of \citet{Kauffmann03}. We exclude GAMA-FIRST matched sources which are identified a spectroscopic-AGN via the BPT diagram (black points). Gold points remain in our sample. Top Right: $WISE$ colour selection of obscured AGN sources. Contours display the GAMA sample, while gold squares display the remaining GAMA-FIRST sample, after BPT rejection of AGN.  We apply a conservative cut in $\mathrm{W3}-\mathrm{W2}<0.125$, red horizontal line, to exclude GAMA-FIRST sources which potentially contain an obscured AGN and also exclude source with $\mathrm{W3}-\mathrm{W2}>-0.5$, red vertical line, as such systems have colours consistent with passive galaxies (and as such their radio emission is unlikely to arise from star-formation.  Bottom left: W4/1.4\,GHz and 1.4\,GHz/K-band AGN selection of \citet{Seymour08}. Gold points show sources that remain in our selection after both the BPT and WISE selection. Bottom right: The NVSS/FIRST 1.4\,GHz flux density ratio as a function of $r$-band effective radius (R$_{\mathrm{e}}$). Gold points display sources which are still in our selection after all previous cuts, while black points display all sources which are detected in both FIRST and NVSS. Note that not all of the remaining same are shown in this panel as only a sub-sample have NVSS detections. We exclude all sources with R$_{\mathrm{e}}$ $>5^{\prime\prime}$ which potentially have resolved out flux in FIRST.}
\label{fig:BPT}
\end{figure*}

A substantial fraction of our GAMA-FIRST sources are likely to be AGN which dominate the 1.4\,GHz number counts at high flux density limits. Given that we aim to produce a robust calibration between radio emission and star-formation, we opt to exclude all sources which potentially have some fraction of their radio emission arising from an AGN and apply multiple cuts to produce a robust, but by design, incomplete sample of star-forming radio galaxies. 1.4\,GHz luminosities for the GAMA-FIRST sample are calculated using the total integrated flux densities (FINT) from the FIRST catalogue, converted to intrinsic luminosity using the GAMA redshifts and k-corrected assuming a power law slope of S$_{\nu}\propto\nu^{-0.7}$ (assuming emission from optically thin synchrotron radiation). For completeness, we also perform our analysis assuming a S$_{\nu}\propto\nu^{-0.6}$ and S$_{\nu}\propto\nu^{-0.8}$ slope and find that it does not significantly change our results. To remove potential AGN-like sources, we apply the following steps:  
\\

$\bullet$ First, we exclude sources which are identified as AGN using the BPT diagnostic \citep{Baldwin81}. We select all GAMA-FIRST galaxies which have [OIII], H${\beta}$, [NII] and H$\alpha$ lines detected at $>2\sigma$. The top left panel of Figure \ref{fig:BPT} displays the distribution of these sources in the BPT diagram. We use the AGN-SF dividing line of \citet{Kauffmann03}, to exclude sources which are identified as AGN via their optical emission line ratios ($i.e.$ we remove all black points in Figure \ref{fig:BPT} from our sample). This removes 236 optically identified AGN.
\\

$\bullet$ This process does not account for heavily obscured (optically thick) AGN, which may not be identified via the BPT method but can still show strong radio emission. In order to remove such sources we apply the \textit{Wide-field Infrared Survey Explorer} ($WISE$) colour selection of obscured AGN in a similar manner to, for example, \cite{Stern12} and \cite{Mateos13}. Figure \ref{fig:BPT} highlights this. The top right panel of Figure \ref{fig:BPT} shows the $WISE$ colours for all GAMA sources (contours) and our GAMA-FIRST matched sample (gold). Here we apply a conservative (more strict than previous works) selection of W1-W2$<$0.125; where W1 and W2 are the observed magnitudes in $WISE$-1 (3.4$\mu$m) and $WISE$-2 (4.6$\mu$m) bands respectively, taken from the GAMA \textsc{lambdar} catalogue (removing 70 sources). 
\\

$\bullet$ We also remove sources which have $WISE$ colours consistent with passive galaxies (as their radio emission is likely to arise from an AGN not SF), using the colours of passive spirals outlined in \cite{Fraser16}, $\mathrm{W3}-\mathrm{W2}>-0.5$, where W3 are the observed magnitudes in $WISE$-3 (12$\mu$m). This removes a further 1277 sources. None of the sources removed here are identified as star-forming using the BPT diagnostic as they have do not have the required BPT emission lines.
\\

\begin{figure*}
\begin{center}
\includegraphics[scale=0.52]{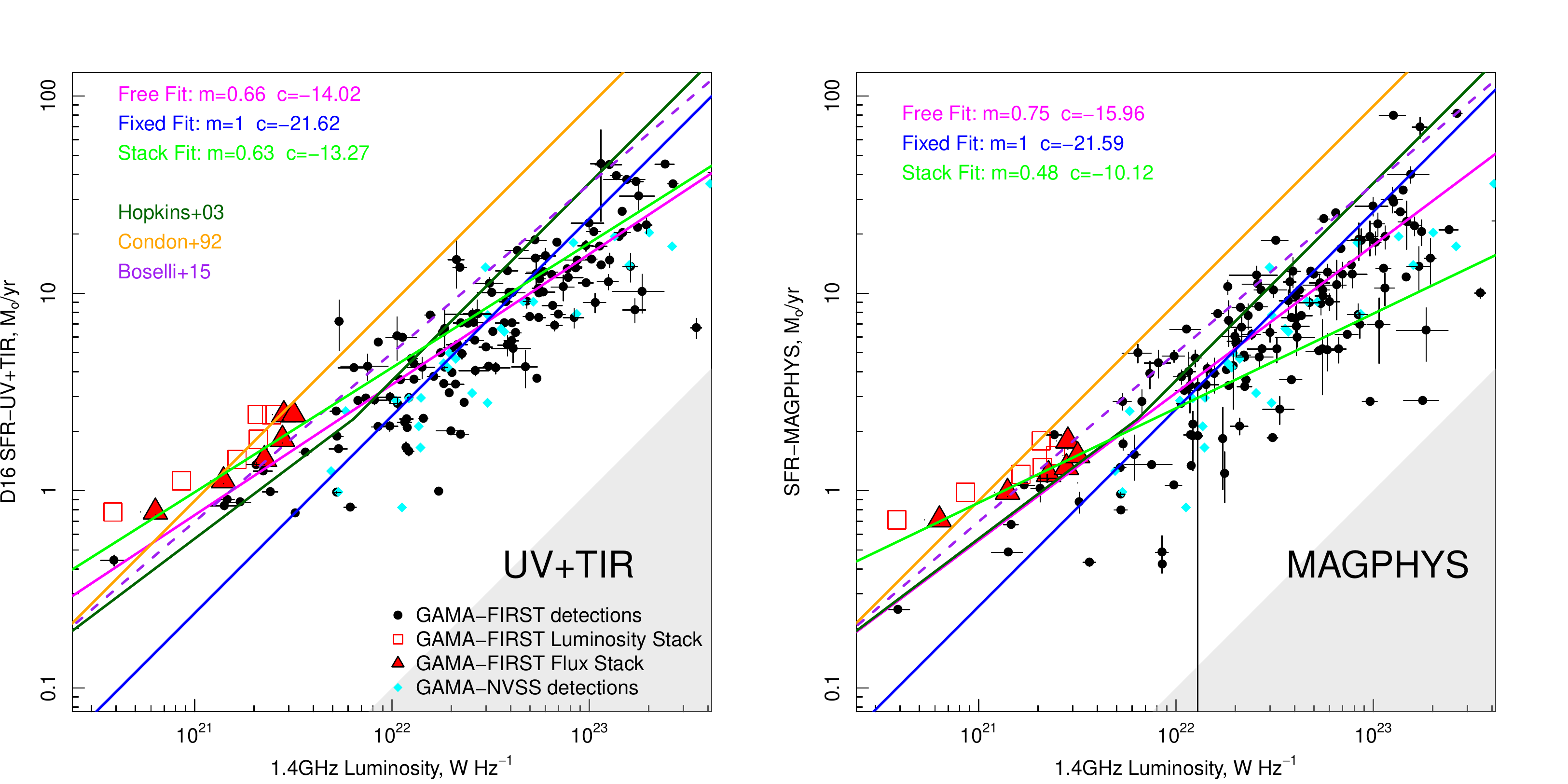}
\end{center}
\caption{Correlation between 1.4\,GHz luminosity and SFR indicators from GAMA outlined in \citet{Davies16b}: UV+Total IR derived SFR (left) and \textsc{magphys} SED-derived SFR (right). Circles display the GAMA-FIRST detected sample. Open squares display our luminosity-stacked samples, while filled triangles display our flux density-stacked samples - where errors are smaller than the plotted points. We fit the relations using \textsc{hyperfit} for both a free slope and normalisation (magenta lines), a fixed m=1 relation (blue lines), a free fit to just the flux density-measured stacked data points (green line). We show the  \citet{Hopkins03}, \citet{Condon92} and \citet{Boselli15} relations as the dark green, orange and purple-dashed lines respectively. For completeness, we also show the small number of sources in our sample with NVSS detections (cyan diamonds) to highlight that including potential `resolved-out' flux in our sample would not significantly change our derived relations. Grey shaded region displays where sources with erroneously high 1.4\,GHz luminosity are excluded. The excluded points fall off this figure and the grey shaded region is only intended to show that we are not biasing our fits by excluding objects in this region.}
\label{fig:calibrations}
\end{figure*}

$\bullet$  We then exclude any source which has a rest-frame 1.4\,GHz luminosity of $>10^{23.5}$W\,Hz$^{-1}$, as such high luminosities may be representative of an AGN (this luminosity would imply SFR$>$200\,M$_{\odot}$\,yr$^{-1}$ using previous calibrations), and also sources with exceedingly large 1.4\,GHz luminosity in comparison to their measured UV+TIR SFR, excluding sources with log$_{10}$[L$_{1.4}$]>log$_{10}$[SFR]+23 (displayed as the grey shaded region in Figure \ref{fig:calibrations}). This selection may remove ultra luminous infrared galaxies (ULIRGS) which potentially have all of their emission arising from star-formation. These sources generally reside at higher redshifts than the GAMA sample however, and thus their potential removal will not affect our derived calibrations. This selection removes a further 170 sources.
\\

$\bullet$  In the bottom left panel of Figure \ref{fig:BPT} we exclude remaining sources which meet the radio-NIR/MIR AGN selection of \cite{Seymour08}. We use a conservative selection to exclude as AGN the 115 sources with log[S$_{22\mu\mathrm{m}}$/S$_{1.4\mathrm{GHz}}$]<0.5 (where S$_{22\mu\mathrm{m}}$ is the \textsc{lambdar} $WISE$-4 (22$\mu$m) flux). This may lead to the removal of low metallicity dwarf galaxies, but this is unlikely to  significantly affect our sample. 
\\

This leaves 172 non-AGN star-forming galaxies in the GAMA-FIRST sample. Using the high resolution FIRST data however, also leads to the possibility of radio flux being `resolved out' for large angular size sources with faint radio emission in their extremities \cite[$i.e.$][]{Jarvis10}. This could potentially lead to an underestimation of source flux density and thus bias any derived calibrations. In order to investigate this we use the the NRAO VLA Sky Survey \citep[NVSS,][]{Condon98}, a 1.4\,GHz survey using the VLA in the more compact D configuration. This compact configuration has poorer resolution than FIRST but greater sensitivity to extended outlying structure. As such, NVSS provides a robust measurement of total 1.4\,GHz flux density, but is more likely to be affected by source confusion. To estimate the fraction of flux that is potentially resolved out, we match to the NVSS catalogue and find 54 sources in our remaining sample have NVSS detections. Figure \ref{fig:BPT} (bottom right) displays the NVSS to FIRST flux density ratio against $r$-band effective radius taken from GAMA. We display both our robust SF sample (gold squares) and all other FIRST-NVSS matches from our initial 1991 sources (black points). Clearly, NVSS measures a larger 1.4\,GHz flux density than FIRST for many sources, but typically finding differences of less then a factor of two. Here we opt to exclude large sources which are most likely to be affected missing flux in FIRST. We do not exclude sources based on their NVSS to FIRST flux density ratio as not all sources have NVSS detections.
\\

$\bullet$ In a similar manner to \cite{Hopkins03} but with a more conservative cut, we exclude all sources from our sample with $r$-band effective radius $>8^{\prime\prime}$, removing a further 24 galaxies. In Figure \ref{fig:calibrations}, displaying our 1.4\,GHz luminosity to SFR relation, we show NVSS measurements for our final sample as cyan points, and highlight that the addition of `resolved out' flux would not strongly affect our derived relations.  
\\

$\bullet$ Last, we then visually inspect all remaining sources for broad H$\alpha$ line emission (potentially broad-line AGN) and/or extended and two component radio emission (potentially lobed radio galaxies), and exclude a further 4 systems.
\\

This leaves a final, highly robust, non-AGN star-forming GAMA-FIRST sample of just 144 galaxies. While this sample is small, we have made every possible effort to exclude any sources of AGN contribution to the radio emission. We present the final GAMA-FIRST sample as the gold squares in Figure \ref{fig:Mz}. This sample largely consists of sources with high radio luminosities and star formation rates (as they are individually detected in the relatively shallow FIRST data). To push to lower radio powers requires the stacking of well-defined populations.

\subsection{GAMA-FIRST Stacking}

To supplement the individually detected GAMA-FIRST galaxies described above, we also perform a stacking analysis of stellar mass selected star-forming galaxies within a volume limited sample from GAMA. 

We use the low contamination and high completeness, volume limited sample of spiral galaxies outlined in Grootes et al (submitted) and D16, and selected following the method presented in \cite{Grootes14} - hereafter GAMA-SPIRALS. Briefly, the sample uses a non-parametric, cell-based, morphological classification algorithm to identify spiral galaxies at $0<z<0.13$. The morphological proxy parameters used in Grootes et al. are the $r$-band effective radius, $i$-band luminosity and single-S\'ersic index \citep[taken from][]{Kelvin12}, importantly avoiding observables which are themselves SFR indicators. We refer the reader to \cite{Grootes14} and Grootes et al. (submitted) for further details.

The red points in Figure \ref{fig:Mz} display the GAMA-SPIRALS sample, which contains 6,366 sources. We then also exclude galaxies which are identified as AGN using the BPT diagnostic, leaving 6,149 sources. This process may still retain heavily obscured AGN, which are problematic to remove from the sample prior to stacking. If included such sources could potentially cause a slight overestimation in the stacked 1.4\,GHz measurements. We do not exclude galaxies that would be identified as AGN using the $WISE$ colour selection in Section \ref{sec:sel} as we wish to keep our GAMA-SPIRALS sample identical to that used in D16, but note that only 21 ($<0.5\%$) sources in our sample would meet the such a selection. 

We split the resulting sample into six stellar mass bins from 9.25$<$log[M$_{*}$/M$_{\odot}$]$<$11.25. We include four intermediate mass bins of $\Delta$log[M$_{*}$/M$_{\odot}$]=0.25, bounded by two larger $\Delta$log[M$_{*}$/M$_{\odot}$]=0.5 bins at the high and low mass end to increase signal to noise in the resultant stacks where either sources are radio faint (the low mass end) or the number density of galaxies is low (the high mass end). Our stacked volumes are displayed as the coloured shaded regions in Figure \ref{fig:Mz}. Stellar mass ranges, median redshifts and number densities of the stacked samples can be found in Table \ref{tab:stacks}.

\begin{table*}
 \caption{Properties of the GAMA-FIRST stacked samples. Column 1: the stellar mass range over which our volume-limited sample of spiral galaxies is stacked. Column 2: the median redshift of the stacked sample. Column 3: the number of sources in the stacked sample. Column 4: the number of individually detected (peak flux density $>$ 0.9mJy) sources in the stack. Column 5: stacked flux density measurement for all sources. Column 6: stacked flux density measurement excluding individually detected sources.  Column 7: Luminosity measurement from flux density-measured stack, using stacked flux density and median redshift. Column 8: Luminosity measurement from luminosity-measured stack, using individual source redshifts. Luminosity measurements are derived from the full stacks only (not excluding detections), but the error range incorporates the difference between the full stack and the stack with detected sources removed. }
 \label{tab:stacks}
 \begin{tabular}{lccccccc}
  \hline
Stellar Mass & Median  & \# & \#  & $S_{1.4}$ - full & $S_{1.4}$ - no detect & L$_{\mathrm{flux-measured}}$ & L$_{\mathrm{lum-measured}}$ \\

 log[M$_{*}$/M$_{\sun}$]  & Redshift  & full & detected & $\mu$Jy$\pm$rms & $\mu$Jy$\pm$rms & $\times10^{21}$ W\,Hz$^{-1}$ & $\times10^{21}$ W\,Hz$^{-1}$\\
 (1) & (2) & (3) & (4) & (5) & (6) & (7) & (8) \\
  \hline
9.25 - 9.75 & 0.100 & 2261 & 7 & 24.2$\pm$5.6 & 24.2$\pm$5.5 & 0.60$\pm$0.14 & 0.39$\pm$0.07   \\
9.75 - 10.00 & 0.107 & 706 & 4 & 46.3$\pm$9.6 & 43.5$\pm$9.7 & 1.33$\pm$0.38 & 0.85$\pm$0.20 \\
10.00 - 10.25 & 0.106 & 565 & 1 & 76.3$\pm$11.1 & 76.3$\pm$11.2 & 2.14$\pm$0.31 & 1.64$\pm$0.22 \\
10.25 - 10.50 & 0.106  & 456 & 3 & 94.1$\pm$12.0 & 93.4$\pm$12.0 & 2.63$\pm$0.34 & 2.11$\pm$0.22 \\
10.50 - 10.75 & 0.108 & 213 & 6 & 91.8$\pm$17.2 & 86.0$\pm$17.5 & 2.67$\pm$0.67 & 2.37$\pm$0.64\\ 
10.75 - 11.25 &  0.111 & 126 & 5 & 100.0$\pm$23.6 & 97.0$\pm$23.7 & 2.98$\pm$0.70 & 2.45$\pm$0.48 \\
  \hline
 \end{tabular}
\end{table*}

We perform the stacking analysis using two different modes both stacking the FIRST data directly, not catalogue measurements. In both modes we apply median stacking to exclude outlying pixels without the need to apply arbitrary cutoffs to the distribution. Median stacking has been found to work successfully when investigating faint sources in FIRST, for example \cite{White07}.  

First, we produce stacks by median combining the pixel values of the FIRST data centred on the positions of the GAMA-SPIRAL samples in each mass bin. We then measure the total integrated flux density at the central beam of the median stack using the \textsc{miriad} \textsc{maxfit} function and derive a 1.4\,GHz luminosity using the median redshift of all sources in the mass bin and k-correcting assuming a power law slope of $\alpha=-0.7$ (median redshifts are given in the second column of Table \ref{tab:stacks}). Hereafter, we will refer to this as the flux density-measured stack. This stacking process essentially assumes that there is no evolution over the redshift range of our sample and that sources are evenly distributed over the redshift range probed.     

Secondly, we determine the individual luminosity of the FIRST data at the position of each of the GAMA-SPIRAL samples. For this we extract a region of the FIRST data centred on the position of the GAMA-SPIRALS source, then convert every pixel value into a luminosity at the source's redshift (again assuming $\alpha=-0.7$). We then median combine the pixel values in each extracted region and again measure the total integrated luminosity at the central beam. Hereafter, we will refer to this as the luminosity-measured stack. This stacked sample uses all distance measurements for individual sources, and hence avoids the assumption of no evolution and even distribution over the redshift range.      

For each stellar mass range we also produce identical stacked samples with the individually detected sources removed. In Table \ref{tab:stacks} we display the median flux density stack measurements for both the full stacks and the stacks with individually detected sources removed. We also display luminosity measurements for both the flux density-measured and luminosity-measured stacks using the full sample. In order to estimate rms errors, we stack the same number of sources as in each stellar mass bin, but at random offset positions in the FIRST data and measure the resultant rms. For the luminosity-measured stack, we calculate the luminosity of all pixels in the offset position using a unique redshift from the GAMA-SPIRALS sample (thus replicating the same redshift distribution in our rms measurements). We do not display luminosity measurements using the stacked sample with individually detected sources removed, but highlight that these only marginally differ from the full stacked sample ($<5\%$). We also include the difference between the full sample and a sample excluding detected sources in our luminosity errors.

In order to avoid including radio emission from sources outside of the GAMA-SPIRALS sample, or repeat stacking within the sample, we confirm that none of our GAMA-SPIRALS sample overlaps with another GAMA source within 5$^{\prime\prime}$, and thus the FIRST beam size. As such, we do not have to exclude potentially confused sources. However, this does not rule out contributions to the emission arising from sources below the GAMA $r$-band selection limit (these sources are likely to be faint in radio emission) or high redshift sources which sit within the beam of the GAMA galaxy. Given it is impossible to remove such sources (as there are no deeper spectroscopic observations in the GAMA regions) we cannot make assessments regarding their contribution to the observed flux density. However, given that the results in the following sections display consistency between our stacked samples and individually detected sources, it is unlikely that such faint galaxies strongly contribute to our derived flux densities. We also do not exclude sources which have $r$-band effective radius$>8^{\prime\prime}$ in our stacked samples, as in the individual detections. While these sources may potentially have `resolved out' flux, we wish to keep the stacked sample identical to that used in D10, and note that an $r$-band effective radius$<8^{\prime\prime}$ cut would only remove 35 sources ($\sim1\%$) from our GAMA-SPIRALS sample. 

\begin{figure*}
\begin{center}
\includegraphics[scale=0.75]{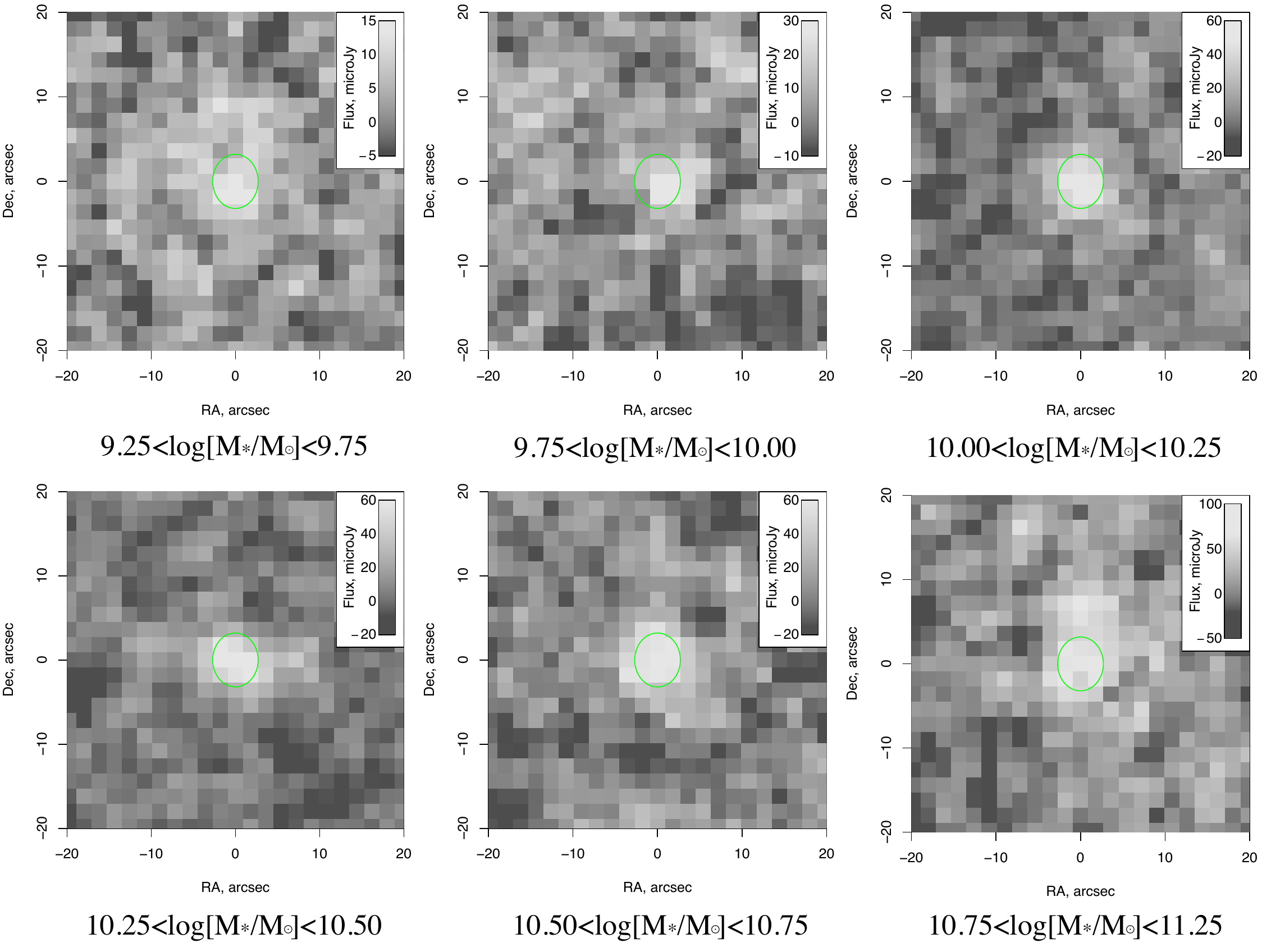}
\end{center}
\caption{Stacked 1.4\,GHz images of our volume limited late-type sample. The green ellipse shows the FIRST beam shape, centred on the stack position. We also produce stacks excluding individually detected sources, and those produced when stacking in luminosity space for each sample, but for clarity we do not show them here; see text for details. }
\label{fig:stacks}
\end{figure*}

Figure \ref{fig:stacks} displays the full stacked GAMA-SPIRAL samples in different stellar mass bins. All stacked values include a multiplication factor of 1.4 to account for ``CLEAN" bias \citep[see][for further details]{White07}. We obtain a $>4.25\sigma$ detection in all bins in our flux density-measured stacks and $>4.5\sigma$ detections in our luminosity-measured stacks .

\section{1.4\,GHz Luminosity-SFR relation}
\label{sec:cals}

Using both the individually detected GAMA-FIRST galaxies and our stacked samples, we investigate the 1.4\,GHz luminosity-SFR relation. D16 provides multiple SFR estimates using 12 different methods for deriving SFR and produces consistent measurement of star-formation across all methods. Here we only compare to the full SED measures of star-formation, UV+TIR (UV+TIR1 in D16) and \textsc{magphys} \citep{daCunha08}. Given the recalibration process in D16, all other GAMA SFR methods will produce similar results to the UV+TIR measurement. We also expect the 1.4\,GHz SFRs to be most closely correlated with long duration measures of star-formation, as they arise from SNe-driven emission. We opt to use the full SED measurements of star-formation over FIR emission only (as has previously been used when calibrating 1.4\,GHz via the FIR-radio relation), as the UV+TIR SFR estimation combines the SF information derived in the FIR with that observed in the UV, and as such is likely to produce a more representative measure of the total star-formation. We also include the \textsc{magphys} SFR as it gives an alternative estimate of the SF, essentially using information from the UV+TIR, but derived using a different fitting method. Both SFR measures used here assume a Chabrier IMF. 

Briefly, the UV+TIR SFR uses the \citet{Brown14} spectrophotometrically calibrated library of galaxy spectra to derive UV and TIR luminosities, from the GAMA 21-band photometry outlined in \cite{Driver16}; using GALEX-UV, SDSS-optical, VIKING-NIR, WISE-MIR and Herschel-ATLAS-FIR data. We follow a Bayesian process, with uniform/uninformative priors on the templates ($i.e.$ each template is assumed to be equally likely). For a particular template, the best fit/maximum likelihood value and the formal uncertainty are analytic (through the usual propagation of uncertainties). The posterior for the best-fit value template is given by marginalising over the full set of templates. By effectively marginalising over template number as a nuisance parameter, we fully propagated the errors, including uncertainties due to template ambiguities.    

\textsc{magphys} SFRs use the \cite{Bruzual03} stellar populations with a \cite{Chabrier03} IMF and assumes an angle-averaged attenuation model of \cite{Charlot00}. This is combined with an empirical NIR-FIR model accounting for PAH features and near-IR continuum emission, emission from hot dust and emission from thermal dust in equilibrium. The code defines a model library over a wide range of star formation histories, metallicities, and dust masses and temperatures, and fits the photometry - forcing energy balance between the observed TIR emission and the obscured flux in the UV-optical. Physical properties (SFR, SFH, metallicity, dust mass, dust temperature) for the galaxy are then estimated from the model fits, giving various percentile ranges for each parameter. Here we use the median SFR$_{0.1Gyr}$ parameter, which provides an estimate for the SFR averaged over the last 0.1\,Gyrs. Errors on SFR$_{\textsc{magphys}}$ are estimated from the 16th-84th percentile range of the SFR$_{0.1Gyr}$ parameter, which encompasses both measurement and fitting errors.

For further details of these SFRs, see the more detailed descriptions in D16. We do not use the favoured radiative transfer-derived SFRs of D16 in this work as we do not have these SFRs for the full GAMA-FIRST sample.

\begin{figure*}
\begin{center}
\includegraphics[scale=0.52]{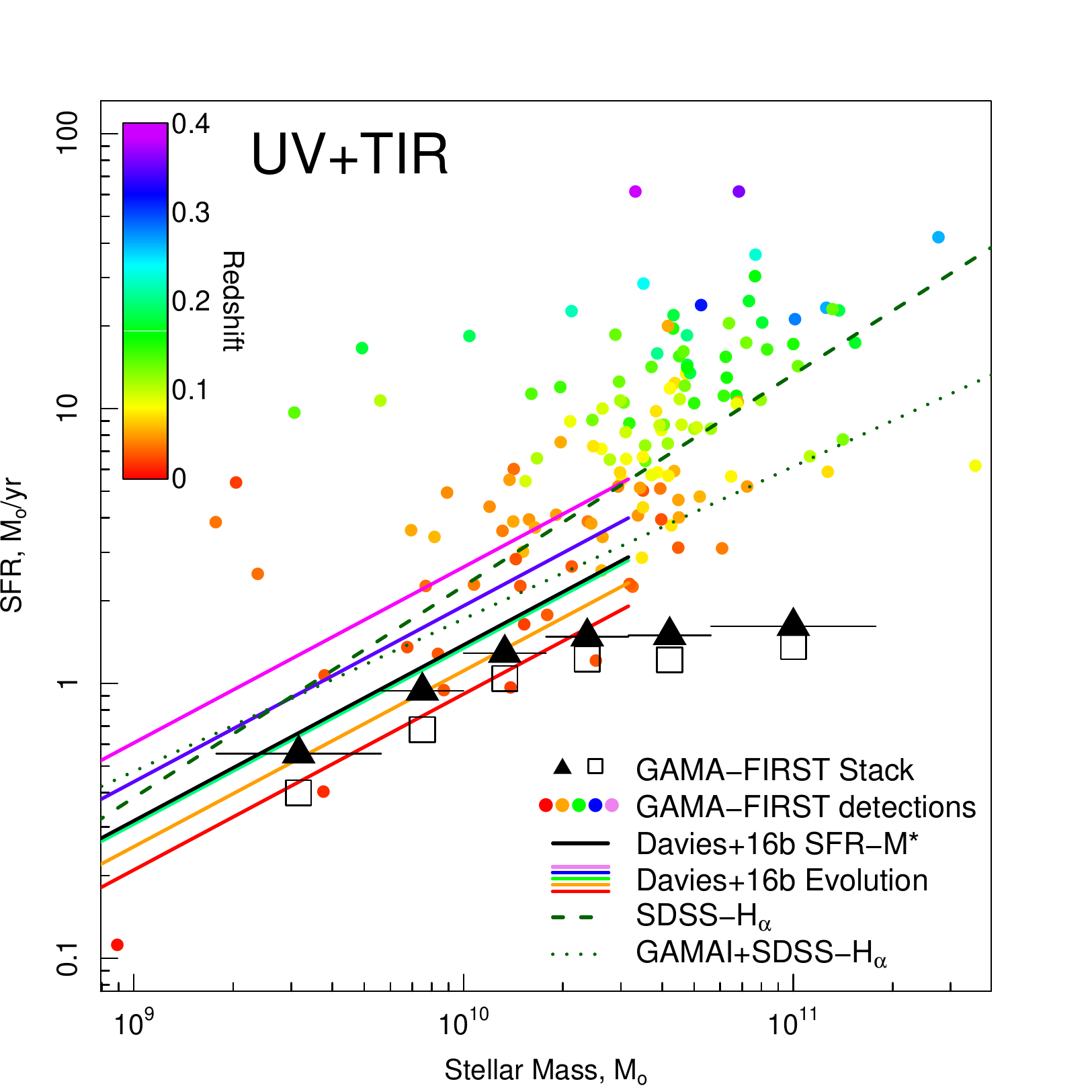}
\includegraphics[scale=0.52]{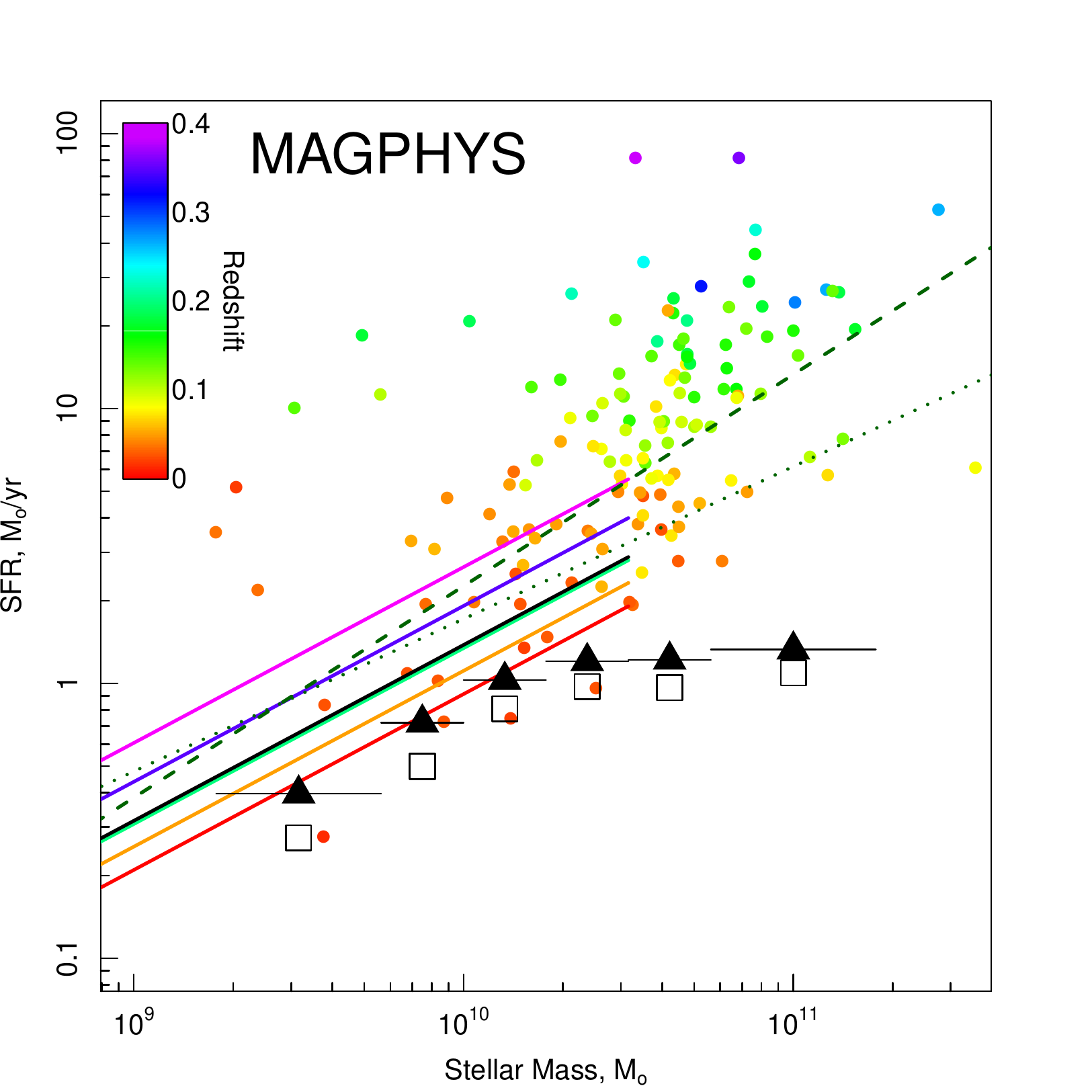}
\end{center}
\caption{The 1.4\,GHz SFR-M$_{*}$ relation in the GAMA regions, derived using using our free-fit luminosity-to-SFR relation for UV+TIR (left) and MAGPHYS (right). Our new 1.4\,GHz-derived SFR-M$_{*}$ relation is consistent with the SFR-M$_{*}$ relation from \citet{Davies16b} at log[M$_{*}$/M$_{\odot}$]<10.5, but tuns over at the high mass end (the known turn over in the SFR-M$_{*}$ relation at high stellar mass). Black triangles and open squares display our flux-measured and luminosity-measured stacked samples respectively, with error bar showing the stacked sample range in stellar mass. Coloured points show the GAMA-FIRST matched sample colour coded by redshift. We also show the SFR-M$_{*}$ fit from \citet{Davies16b} scaled to various redshifts, given the normalisation evolution taken from Eq 20 of D16 and colour coded on the same redshift scale as the data points. The black line displays the direct SFR-M$_{*}$ fit from D16 to an identical sample used in our stacking analysis here. Consequently, the stacked data points should be directly compared to the black line. We also show the H$\alpha$-derived SFR-M$_{*}$ fits from SDSS at z=0 \citep{Elbaz07} and GAMA I + SDSS at $z<0.1$ \citep{Lara-lopez13} as the green dashed and dotted lines respectively. Errors in median SFR (including fitting errors) are smaller than the symbols.}
\label{fig:KE12}
\end{figure*}

Figure \ref{fig:calibrations} displays the 1.4\,GHz Luminosity-SFR relation for both the UV+TIR SFR and \textsc{magphys} SFRs from D16. Individually detected sources from the GAMA-FIRST sample are displayed as circles while the flux density-measured and luminosity-measured stacks are displayed as filled triangles and open squares respectively. Both methods for determining the stacked fluxes are found to be within the scatter of the individually detected sources, suggesting that stacking method does not strongly affect our results. We do find that the luminosity-measured stacks produce systematically lower luminosity measurements, which potentially suggests that in using a flux stack and median redshift over-predicts to true luminosity. This may be due to the fact that the sources are not uniformly distributed over the redshift of our stacked sample.

We show previously published relations outlined in \cite{Hopkins03} (from SDSS-FIRST), \cite{Boselli15} (from the $Herschel$ Reference Survey, K-band selected sample) and \cite{Condon92}, as the dark green, purple-dashed and orange solid lines respectively. The \cite{Hopkins03} line is plotted as a broken power law to account for the scaling for non-thermal radio continuum emission from dwarf galaxies, applied in their relation. All relations are scaled to a Chabrier IMF using the conversions outlined in \cite{Haarsma00}, for Miller-Scalo to Salpeter, and \cite{Driver13}, for Salpeter to Chabrier. 

We then fit the 1.4\,GHz luminosity-SFR relation linearly in a number of ways using the multi-dimensional MCMC fitting \textsc{[r]} package \textsc{hyperfit}\footnote{http://hyperfit.icrar.org/} \citep{Robotham15}. Firstly, we fit the full distribution using a fixed, m=1, slope (blue line), these fits are almost identical for both the UV+TIR and \textsc{magphys} SFRs but have an offset normalisation from the \cite{Hopkins03} and \cite{Condon92} relations. Secondly we fit the distributions with a free slope and normalisation (magenta line), these fits have a slightly different slope between the UV+TIR and \textsc{magphys} SFRs. Interestingly for both the UV+TIR and \textsc{magphys} SFRs this fit has a similar slope and normalization to the lower 1.4\,GHz broken power-law component, for dwarf galaxies, of the \cite{Bell03} and \cite{Hopkins03} relation ($i.e.$ the dark green and magenta fits have a similar slope at $L_{\mathrm{1.4GHz}}<6.4\times10^{21}$\,W\,Hz$^{-1}$). Lastly, we fit the distributions using just the flux density-measured stacks (green line).

All fits take the form of:
\begin{equation}
\mathrm{log}_{10}[SFR\mathrm{(M_{\odot}yr^{-1})}]=m \times \mathrm{log}_{10}[L\mathrm{_{1.4GHz}(W\,Hz^{-1})}]+C
\end{equation}
\noindent with parameters, $m$ and $C$, given in the figure. Given our free fit (which are the best fit to the full dataset) we suggest a new calibration to the 1.4\,GHz-SFR relation as:

\begin{equation}
\label{eq:UVTIR}
\mathrm{log}_{10}[SFR_{\mathrm{UV+TIR}}]=0.66\pm0.02 \times \mathrm{log}_{10}[L_{\mathrm{1.4}}]-14.02\pm0.39
\end{equation}

\begin{equation}
\label{eq:MAGP}
\mathrm{log}_{10}[SFR_{\mathrm{MAGP}}]=0.75\pm0.03 \times \mathrm{log}_{10}[L_{\mathrm{1.4}}]-15.96\pm0.58
\end{equation}

Interestingly, we find best fit relations with sub-linear slopes ($i.e.$ $m\ne1$). Given that thermal radio emission scales linearly with SFR (from fundamental theory of the emission processes), this must mean that the non-thermal component is sub-linear. This is consistent with non-calorimetric models of non-thermal emission in galaxy disks \citep[$e.g.$][]{Niklas97,Bell03, Lacki10, Irwin13, Basu15}, where cosmic ray electrons do not lose all of their energy before escaping galaxies and not all of their energy is radiated as synchrotron radio emission. These models predict a SFR\,$\propto$\,$L\mathrm{_{1.4}}^{0.73-0.9}$ relation  (consistent slope of our \textsc{magphys} fits). However, the somewhat extreme non-calorimetric model are seemingly in conflict with the tightness of the far-IR-radio relation over a broad range of physical properties of the host galaxy \citep[$e.g$ see diecussion in][]{Lacki10}. While linear, calorimetric, fits ($m=1$, blue lines) are not in strong conflict with our data (specifically for the \textsc{magphys} relations), non-calorimetric models for radio emission will require further investigation in the MeerKAT/ASKAP/SKA era.

\section{The 1.4\,GHz SFR-M$_{*}$ Relation}

Using the 1.4\,GHz luminosity-SFR calibration derived above, it is possible to explore the 1.4\,GHz  SFR-M$_{*}$ relation (Figure \ref{fig:KE12}). We display the flux-measured stacked data points as solid black triangles, luminosity-measured stacked data points as open squares and the individually detected GAMA-FIRST sample are shown as circles colour coded by their redshift. We show the SFR-M$_{*}$ relation fit for the GAMA-SPIRALS sample using the radiative transfer SFRs from D16 as the black solid line, and the same fit at various redshifts (colour coded in the same manner as the data points) using the evolution of the normalisation of the SFR-M$_{*}$ relation using Eq 20 of D16.   Green dashed and dotted lines show the H$\alpha$-derived SFR-M$_{*}$ fits from SDSS at z=0 \citep{Elbaz07} and GAMA I + SDSS at $z<0.1$ \citep{Lara-lopez13} respectively. These do not include the turnover at high stellar masses as they are fit linearly. 

We find that the slope and normalisation of the 1.4\,GHz SFR-M$_{*}$ relation from our stacked samples, using our new calibration (black triangles), has the same slope to that derived in D16 at log[M$_{*}$/M$_{\odot}$]<10.5 (Figure \ref{fig:KE12}); the black line in this figure is the fit using the same sample that is stacked in this work, but with slight normalisation offset ($\sim$0.05\,dex for the flux-weighted stacks using UV+TIR). While the \textsc{magphys} stacked data points are $\sim$0.3\,dex lower than the D16 relation, this is roughly consistent with the offset in normalisation between the UV+TIR and \textsc{magphys} SFR-M$_{*}$ relations in Figure 8 of D16. 

The slope of the 1.4\,GHz SFR-M$_{*}$ relation flattens at log[M$_{*}$/M$_{\odot}$]>10.5. This is expected given the well known turn over in the SFR-M$_{*}$ relation at high stellar masses \citep[see][and discussion in D16]{Whitaker14,Johnston15,Lee15,Schreiber15,Gavazzi15, Tomczak16}. The turnover observed here is severe however, given that our stacked sample is based on purely spiral galaxies. We do highlight that there is a turn over observed in other SFR indicators using the same sample \citep[see coloured circles in Figure 8 of][]{Davies16b}, but this is less extreme (although only measured to log[M$_{*}$/M$_{\odot}$]=10.5). Potentially we are simply observing the increasing contribution of passive bulges, in terms of specific SFR, in galaxies at the high mass end. Further studies into the high mass turn over in comparison to galaxy morphology and components will be the subject of upcoming work (Davies et al, in prep).     

Primarily the GAMA-FIRST individually detected galaxies lie well above the SFR-M$_{*}$ relation at their redshift, suggesting they are star-bursting galaxies. This is unsurprising given that they are detected in the relatively shallow FIRST data. The exceptions to this are the very local galaxies (red points), which are mostly consistent with the SFR-M$_{*}$ relation; very nearby sources can be detected by FIRST to lower SFRs. 

It is also interesting to note that using the individually detected GAMA-FIRST galaxies, one would not have been able to define the 1.4\,GHz SFR-M$_{*}$ relation given the small number of sources spread over a large redshift range. This highlights the power in performing optically-motivated source stacking of radio continuum data using surveys such as GAMA. The stacked data points allow us explore the 1.4\,GHz SFR-M$_{*}$ relation to lower stellar masses than those probed by the individual detected sources and for the first time, show that the slope and normalisation of the 1.4\,GHz SFR-M$_{*}$ relation over 2 decades in stellar mass is consistent with previous estimates using other multiple SFR tracers.

\section{Predictions for GAMA-ASKAP}

Despite the recent advancements in studying 1.4\,GHz emission from galaxies in large area surveys, the relatively shallow depth of current radio continuum surveys such as FIRST and NVSS, and the small area of deep radio continuum surveys, such as VLA-COSMOS \citep{Schinnerer07} and ATLAS 1.4\,GHz \citep{Hales14}, have limited the number of sources with detectable 1.4\,GHz continuum emission with which to derive SFRs. This is set to change dramatically with the advent of new deep large area continuum surveys from the Square Kilometre Array (SKA) and its precursors such as  ASKAP-EMU \citep{Norris11} and MeerKAT-MIGHTEE \citep{Jarvis12}. One of the key scientific goals of the SKA is to measure the cosmic star-formation history using the radio continuum as a dust-unbiased tracer of star-formation \citep[see][]{Ciliegi15,Jarvis15a,Jarvis15b}.

A potential limiting factor in the use of the 1.4\,GHz SFR tracer in large area surveys however, is the lack of robust spectroscopic redshifts, with which to derive 1.4\,GHz luminosities from observed flux densities and aid in the separation of AGN/SF-like sources. EMU is likely to detect 70 million galaxies, of which only a small faction will have spectroscopic redshifts, mostly at low-$z$ ($z<0.25$) from EMUs sibling HI spectral line survey WALLABY \citep[see][]{Koribalski12} and the local galaxy redshift survey, Taipan. Beyond the very local Universe, EMU will have to either rely on photometric redshifts, undertake additional spectroscopic observations, or use redshifts from existing large area surveys. 

The GAMA survey and upcoming Wide Area VISTA Survey \citep[WAVES,][]{Driver16b}, are ideally suited to providing a large number of spectroscopic redshifts. GAMA contains redshifts for $\sim$280,000 galaxies in the EMU footprint at $z<0.4$. In addition, GAMA provides an extensive database of multi-wavelength observations and value added catalogues of FIR luminosities, stellar masses, dust masses, metallicities, environmental metrics and most importantly, multiple metrics of star formation with which to compare to the observed EMU luminosities (see D16). The upcoming WAVES survey will add $\sim2$M galaxies to this sample to $z<0.8$, which will be invaluable in providing redshifts, environmental metrics, and derived parameters for EMU sources. The combination of GAMA/WAVES with the ASKAP surveys \citep[EMU, WALLABY and Deep Investigations of Neutral Gas Origins, DINGO,][]{Meyer09} will produce a formidable dataset with which to study galaxy evolution over an extensive redshift baseline.   

Using the 1.4\,GHz luminosity to SFR relations derived in the previous section we make predictions for the number of GAMA star-forming galaxies that are likely to be detected in upcoming deep radio continuum surveys using ASKAP.

We take the full GAMA II$_{Eq}$ SFRs derived in D16 for a number of different SFR methods, and use the 1.4\,GHz luminosity to SFR relation to predict the rest-frame 1.4\,GHz luminosity for all GAMA II$_{Eq}$ sources. Assuming $\alpha=-0.7$ and the GAMA redshift, we then convert each luminosity to a predicted observed flux density. We then exclude all sources which are detected as an AGN using the BPT diagnostic or have $WISE$ colours consistent with an AGN ($\mathrm{W1}-\mathrm{W2}>0.125$ - as in the top two panels of Figure \ref{fig:BPT}). We also exclude all sources which do not have a $>2\sigma$ detection in the observable used to determine the source's SFR; such sources may have erroneous measurements of star formation. For \textsc{magphys} we only consider sources where the derived SFR is greater then twice the error. 

\begin{figure*}
\begin{center}
\includegraphics[scale=0.65]{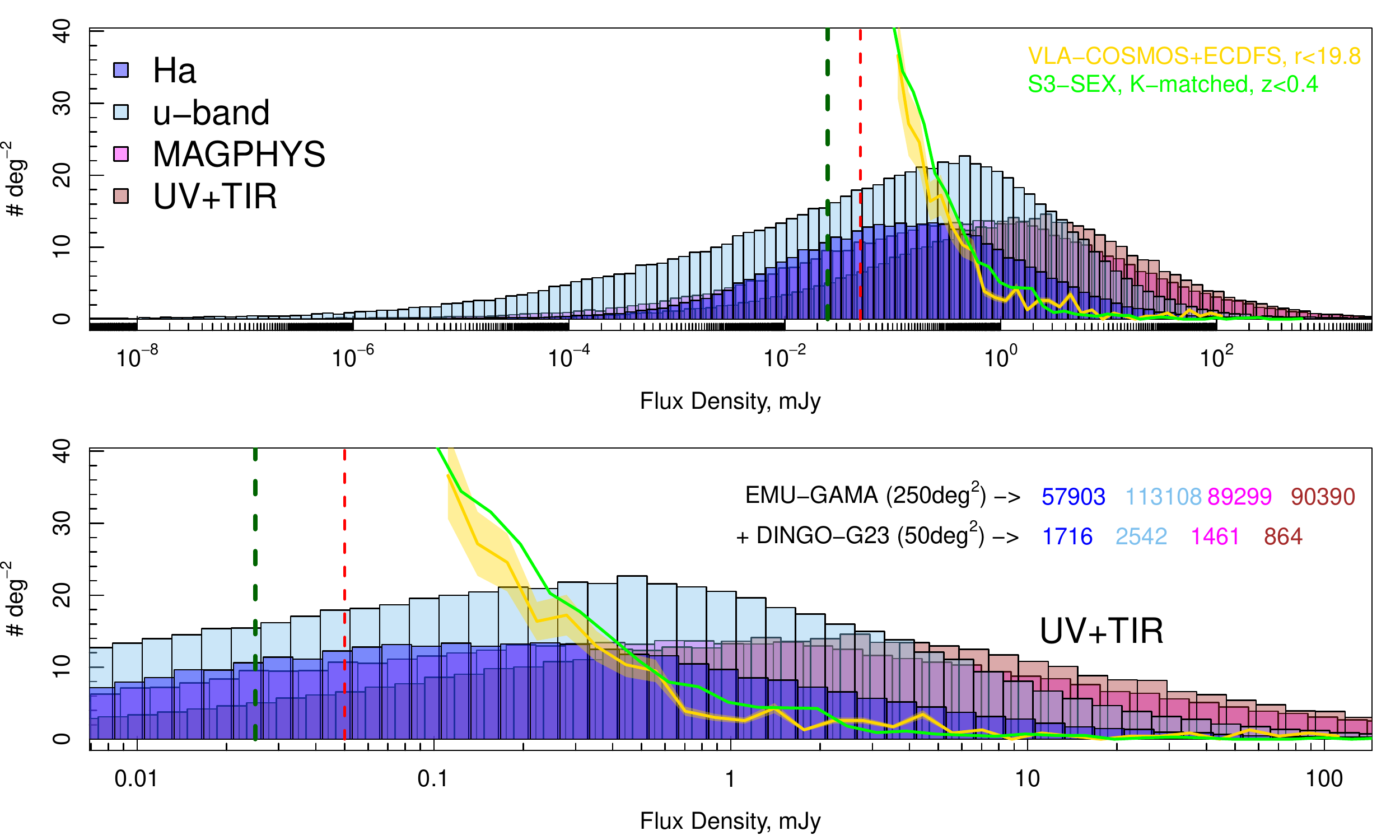}
\includegraphics[scale=0.65]{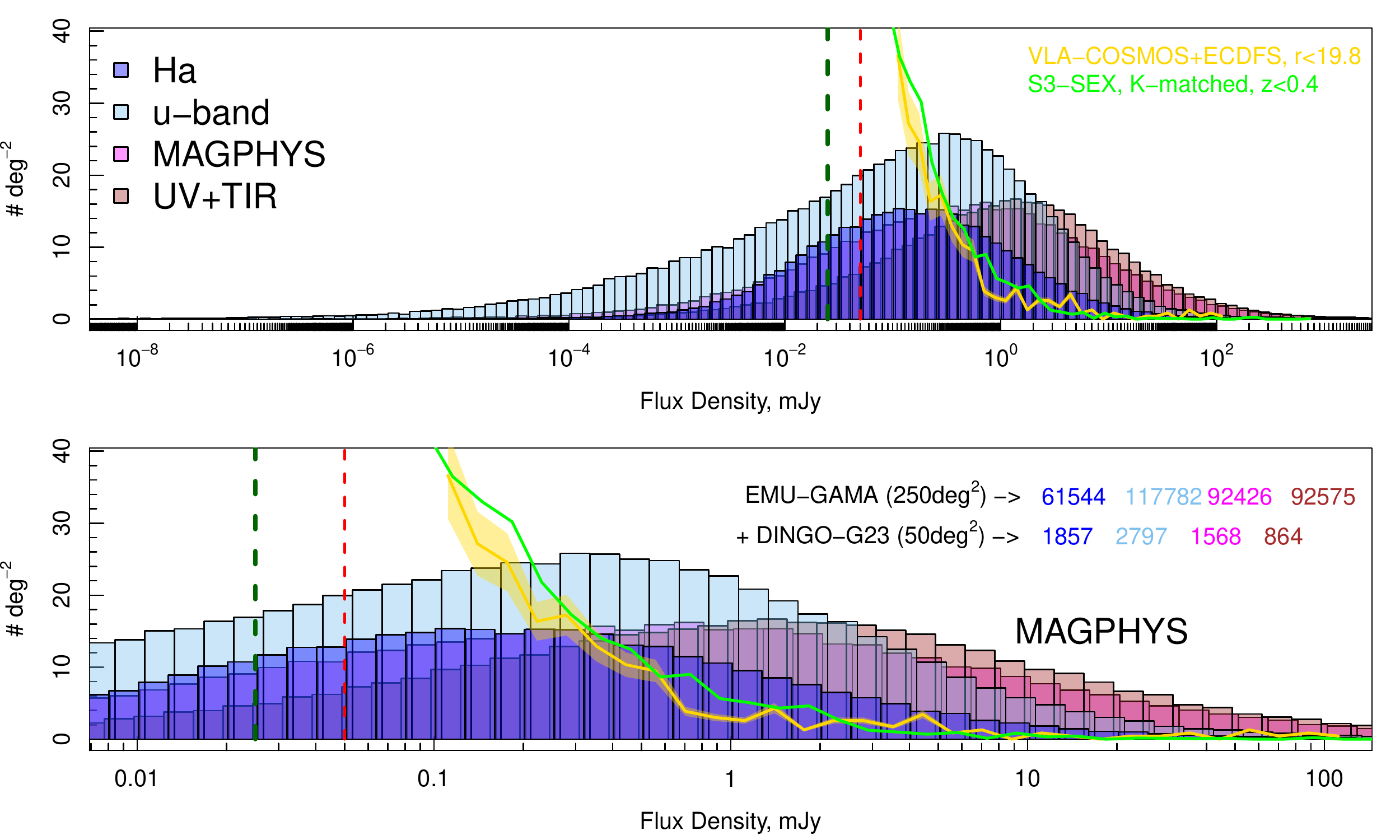}
\end{center}
\caption{The predicted 1.4\,GHz flux density distribution of GAMA star-forming galaxies using different SFR measures. We take the observed SFR from GAMA, convert to a 1.4\,GHz luminosity using the UV+TIR (top two panels) and MAGPHYS (bottom two panels) free-fit relation, and scale to an observed flux density using the sources' true redshift. The dashed vertical lines display the ASKAP-EMU (red) and ASKAP-DINGO (green) 1.4\,GHz 5$\times$rms continuum limits. The top panel shows the full distribution and the bottom panel shows a zoomed in region at the high flux density end. Using these values we can make predictions for the total number of GAMA star forming galaxies that are likely to be detected in EMU and DINGO (given in the bottom panel for each SFR tracer). The gold line displays the 1.4\,GHz number density of $r$-mag<19.8 sources from the VLA-COSMOS and VLA-ECDFS surveys, with error band estimated for the 16\% cosmic variance error. The green line displays the prediction from a GAMA-like sample in S$^{3}$-SEX.}
\label{fig:predict}
\end{figure*}

Figure \ref{fig:predict} displays the predicted distribution of 1.4\,GHz flux densities from all GAMA II$_{Eq}$ sources using four of the different SFR methods discussed in D16, for both UV+TIR (top) and \textsc{magphys} (bottom) calibrations. H$\alpha$, $u$-band and UV+TIR SFRs are all derived using the recalibration process detailed in D16. The various histograms in each figure highlight different predictions for the 1.4\,GHz flux density distribution from GAMA sources, assuming different SFR measures in GAMA ($i.e.$ how does the choice of SFR tracer in GAMA affect the prediction), while the two sets of panels show the variation based on the 1.4\,GHz to SFR calibration used (either Eq. \ref{eq:UVTIR}, using UV+TIR, or Eq. \ref{eq:MAGP}, using \textsc{magphys}). A potential caveat of this analysis is that the 1.4\,GHz calibration derived in this work is for late-type star-forming galaxies, but this calibration is applied to all GAMA galaxies (and may not be appropriate in all cases).

We compare this predicted distribution to the observed number density of sources in deep radio continuum surveys, using a combination of the VLA-COSMOS \citep{Schinnerer04} and VLA-ECDFS \citep{Miller08} surveys. First, we use the VLA-COSMOS 1.4\,GHz catalogue of \cite{Schinnerer07}. In order to produce a representative sample of GAMA-like galaxies, we perform a 3$^{\prime\prime}$ position match (to be consistent with our previous matching) of the VLA-COSMOS catalogue to the COSMOS photometric catalogue of \cite{Capak07} and retain matches which have $r<19.8$, the GAMA selection limit. We then combine this with the VLA-ECDFS optical counterpart catalogue of \cite{Bonzini2012}, once again cut at $r<19.8$. The gold line in Figure \ref{fig:predict} displays the number density as a function of flux density for $r<19.8$ sources in the combined VLA-COSMOS+ECDFS. We also include a 16\% cosmic variance error (gold band), calculated using the prescription in \cite{Driver10} \footnote{See http://cosmocalc.icrar.org/} to account for the small volume coverage of these deep surveys at low-$z$. However, this process does not account for these deep surveys resolving out flux for low redshift sources. 

In addition, we display the predicted number density of sources from the SKA Simulated Skies (S$^{3}$) simulations of extragalactic radio continuum sources (S$^{3}$-SEX) outlined in \cite{Wilman08}. To make this comparable to a potential GAMA-FIRST sample, we take all sources from S$^{3}$-SEX at $z<0.4$ and match them to the the observed distribution of GAMA sources in K-band magnitude (in the absence of stellar mass in the S$^{3}$-SEX catalogues). We take the observed K-band distribution from GAMA and randomly sample from the S$^{3}$-SEX simulated sources at $z<0.4$ to produce the same K-mag number density distribution. While this is not ideal, and should be treated as a very loose prediction for a GAMA-like sample, it aims to produce as close to a GAMA-representative sample from S$^{3}$-SEX as possible. We show the number density of these S$^{3}$-SEX sources as the green line in Figure \ref{fig:predict}       

Strikingly the predicted number density using the H$\alpha$ SFRs in GAMA is very close to the observed distribution from VLA-COSMOS+ECDFS and the predicted distribution from S$^{3}$-SEX. This suggests that our predictions are producing a comparable number density of 1.4\,GHz sources to the observed distribution at $>0.3$mJy. The H$\alpha$ SFR appears most well correlated with the observed distribution, potentially as both mechanisms probe the central regions of galaxies, particularly when using the high resolution FIRST imaging. It is also important to remember that this H$\alpha$ SFR has been previously re-calibrated using the radiative transfer-derived SFR in D16. 

Using these distributions it is therefore possible to make predictions for the number of GAMA sources which are likely to be detectable in EMU and DINGO-continuum. The green and red dashed lines in Figure \ref{fig:predict} show the 5$\times$rms limits of EMU and DINGO-continuum, taken as 0.05mJy and 0.025mJy respectively. Taking the number density of GAMA sources above the EMU limit and scaling to the full GAMA volume ($\sim250$\,deg$^{2}$) we obtain the predicted number of GAMA-EMU sources, for each SFR method, given in the top right corner of the figure. We then also predict the number of sources which will be undetected in EMU but detectable in the DINGO-continuum overlap with the GAMA-G23 field ($\sim50$deg$^{2}$). The predictions range from $\sim$65,000 to $\sim$115,000 GAMA-ASKAP sources depending on SFR tracer used. This suggests that $\sim25-45\%$ of GAMA star forming galaxies (plus many radio-loud AGN) are likely to be detected in GAMA-EMU.

The large uncertainty in these predictions shows the differences/confusion in deriving SFRs using multiple tracers with different assumptions and sources of error, and once again highlights the need for a robust dust unbiased tracer of star-formation. A comparison between the true distribution of star-forming galaxies in GAMA-ASKAP and these predictions, will help constrain the robustness of individual SFR tracers.

Clearly, using current surveys to investigate the dust-unbiased evolution of star-formation in the local Universe is limited by the depth of large area radio surveys - the GAMA-FIRST sample is heavily constrained by the number of FIRST detections. However, with the advent of the ASKAP surveys we will no longer be constrained by the lack of radio detections, but in fact by the number of robust redshifts available to match to secure radio sources. This highlights the necessity for further deep, wide area spectroscopic surveys such as WAVES. Applying the same prescription as above to the current WAVES mock catalogues, we can predict that $\sim500,000$ of the $\sim2$\,million WAVES sources are likely to be detected by EMU - producing an impressive dataset with which to study galaxy evolution from the NUV through to the radio continuum.

\section{Conclusions}

We have defined a robust sample of individually detected GAMA-FIRST galaxies and produced a stellar mass weighted stack in the FIRST images at the position of a volume limited spirals sample in GAMA. We exclude AGN from our sample using the BPT diagram, radio power, $WISE$ colours, 1.4\,GHz-W4 relation and $r$-band size to produce an uncontaminated star-forming galaxy sample. We then compare the 1.4\,GHz luminosity of our sample to previously derived SFRs from GAMA and derive new 1.4\,GHz luminosity to SFR calibrations. We derive the dust-unbiased SFR-M$_{*}$ relation to show that our new calibrations produce a relation with a same slope and normalisation roughly consistent to that previously derived for GAMA using 12 other SFR methods \cite{Davies16b}, highlighting the power of optically motivated source stacking in large area radio surveys. This also shows that our calibrations are robust in deriving SFRs from radio luminosity. We do find a significant turn over at the high mass end, potentially highlighting a true turnover in the distribution, which is not observed in D16 (although they do not probe as high in stellar mass). 

We use this relation to make predictions for the number of GAMA sources that are likely to be detected in radio continuum by upcoming ASKAP surveys. Using the GAMA H$\alpha$ SFRs we obtain a prediction which is consistent with existing deep radio surveys at flux density $>0.1$\,mJy. We predict that between 65,000 and 115,000 GAMA sources ($\sim$25-45\%) are likely to be detected by ASKAP, and in the near future a further $\sim500,000$ EMU sources will have spectroscopic redshifts from WAVES. The combination of deep, large area radio surveys and spectroscopic redshift surveys will revolutionise our view of dust-unbiased star-formation in the Universe.

\section*{Acknowledgements}

GAMA is a joint European-Australasian project based around a spectroscopic campaign using the Anglo-Australian Telescope. The GAMA input catalogue is based on data taken from the Sloan Digital Sky Survey and the UKIRT Infrared Deep Sky Survey. Complementary imaging of the GAMA regions is being obtained by a number of independent survey programs including GALEX MIS, VST KiDS, VISTA VIKING, WISE, Herschel-ATLAS, GMRT and ASKAP providing UV to radio coverage. GAMA is funded by the STFC (UK), the ARC (Australia), the AAO, and the participating institutions. The GAMA website is http://www.gama-survey.org/.

\bsp	
\label{lastpage}
\end{document}